\documentclass[superscriptaddress,aps,preprintnumbers,amsmath,showpacs,amssymb,prd,nofootinbib,reprint]{revtex4-1}
\usepackage{bm, color} 
\usepackage{amssymb,amsfonts,slashed,amsthm,amsmath,graphicx, soul}
\usepackage{xcolor}
\usepackage[caption=false]{subfig}
\usepackage{multirow}
\usepackage[normalem]{ulem}
\begin{document}


\renewcommand{\figurename}{Fig.}
\renewcommand{\tablename}{Table.}
\newcommand{\Slash}[1]{{\ooalign{\hfil#1\hfil\crcr\raise.167ex\hbox{/}}}}
\newcommand{\bra}[1]{ \langle {#1} | }
\newcommand{\ket}[1]{ | {#1} \rangle }
\newcommand{\bef}{\begin{figure}}  \newcommand{\eef}{\end{figure}}
\newcommand{\bec}{\begin{center}}  \newcommand{\eec}{\end{center}}
\newcommand{\laq}[1]{\label{eq:#1}}  
\newcommand{\dd}[1]{{d \o d{#1}}}
\newcommand{\Eq}[1]{Eq.~(\ref{eq:#1})}
\newcommand{\Eqs}[1]{Eqs.~(\ref{eq:#1})}
\newcommand{\eq}[1]{(\ref{eq:#1})}
\newcommand{\Sec}[1]{Sec.\ref{chap:#1}}
\newcommand{\ab}[1]{\left|{#1}\right|}
\newcommand{\vev}[1]{ \left\langle {#1} \right\rangle }
\newcommand{\bs}[1]{ {\boldsymbol {#1}} }
\newcommand{\lac}[1]{\label{chap:#1}}
\newcommand{\SU}[1]{{\rm SU{#1} } }
\newcommand{\SO}[1]{{\rm SO{#1}} }
\def\({\left(}
\def\){\right)}
\def\dt{{d \o dt}}
\def\diag{\mathop{\rm diag}\nolimits}
\def\Spin{\mathop{\rm Spin}}
\def\O{\mathcal{O}}
\def\U{\mathop{\rm U}}
\def\Sp{\mathop{\rm Sp}}
\def\SL{\mathop{\rm SL}}
\def\tr{\mathop{\rm tr}}
\def\ebq{\end{equation} \begin{equation}}
\newcommand{\OR}{~{\rm or}~}
\newcommand{\AND}{~{\rm and}~}
\newcommand{\EV}{ {\rm \, eV} }
\newcommand{\KEV}{ {\rm \, keV} }
\newcommand{\MEV}{ {\rm \, MeV} }
\newcommand{\GEV}{ {\rm \, GeV} }
\newcommand{\TEV}{ {\rm \, TeV} }
\def\o{\over}
\def\a{\alpha}
\def\b{\beta}
\def\c{\varepsilon}
\def\d{\delta}
\def\e{\epsilon}
\def\f{\phi}
\def\g{\gamma}
\def\h{\theta}
\def\k{\kappa}
\def\l{\lambda}
\def\m{\mu}
\def\n{\nu}
\def\p{\psi}
\def\q{\partial}
\def\r{\rho}
\def\s{\sigma}
\def\t{\tau}
\def\u{\upsilon}
\def\w{\omega}
\def\x{\xi}
\def\y{\eta}
\def\z{\zeta}
\def\D{\Delta}
\def\G{\Gamma}
\def\H{\Theta}
\def\L{\Lambda}
\def\F{\Phi}
\def\P{\Psi}
\def\S{\Sigma}
\def\me{\mathrm e}
\def\ol{\overline}
\def\tl{\tilde}
\def\*{\dagger}
%


\newcommand{\wy}[1]{\textcolor{blue}{#1}}
\newcommand{\WY}[1]{\textcolor{blue}{\sffamily #1}}
\newcommand{\WYC}[1]{\textcolor{blue}{WY:\bf  \sffamily #1}}
\newcommand{\JL}[1]{\textcolor{orange}{\sffamily #1}}
\newcommand{\JLC}[1]{\textcolor{orange}{JL:\bf  \sffamily #1}}
\newcommand{\km}[1]{\textcolor{magenta}{#1}}
\newcommand{\KM}[1]{\textcolor{magenta}{\sffamily #1}}
\newcommand{\KMC}[1]{\textcolor{magenta}{[KM: {\bf  \sffamily #1}]}}
\newcommand{\FT}[1]{\textcolor{red}{\sffamily #1}}
\newcommand{\FTC}[1]{\textcolor{red}{[FT]:\bf  \sffamily #1}}

\preprint{TU-1198}


\title{ 
Gravitational Waves from Domain Wall Collapse, and Application to Nanohertz Signals with QCD-coupled Axions
}

\author{
Naoya Kitajima
}
\affiliation{Frontier Research Institute for Interdisciplinary Sciences, Tohoku University, Sendai, Miyagi 980-8578, Japan}
\affiliation{Department of Physics, Tohoku University, 
Sendai, Miyagi 980-8578, Japan} 

\author{
Junseok Lee
}

\affiliation{Department of Physics, Tohoku University, 
Sendai, Miyagi 980-8578, Japan} 

\author{
Kai Murai
}

\affiliation{Department of Physics, Tohoku University, 
Sendai, Miyagi 980-8578, Japan}

\author{
Fuminobu Takahashi
}
\affiliation{Department of Physics, Tohoku University, 
Sendai, Miyagi 980-8578, Japan}

\author{
Wen Yin
}
\affiliation{Department of Physics, Tohoku University, 
Sendai, Miyagi 980-8578, Japan}

\begin{abstract}
{We study for the first time the gravitational waves generated during the collapse of domain walls, incorporating the potential bias in the lattice simulations. The final stages of domain wall collapse are crucial for the production of gravitational waves, but have remained unexplored due to computational difficulties. As a significant application of this new result, we show that the observed NANOGrav, EPTA, PPTA, and CPTA data, which indicate stochastic gravitational waves in the nanohertz regime, can be attributed to axion domain walls coupled to QCD. In our model, non-perturbative effects of QCD induce a temperature-dependent bias around the QCD crossover, inducing the rapid collapse of the domain walls. We use sophisticated lattice simulations that account for the temperature-dependent bias to measure the gravitational waves resulting from the domain wall annihilation. We also discuss the future prospects for accelerator-based searches for the axion and the potential for the formation and detection of primordial black holes.
}
\end{abstract}


\maketitle
\flushbottom

\vspace{1cm}

{\bf Introduction.--}
There are various symmetries in physics beyond the Standard Model, and their spontaneous breaking during the evolution of the universe generates various topological defects~\cite{Zeldovich:1974uw,Kibble:1976sj,Vilenkin:1981zs} (see Ref.~\cite{Vilenkin:2000jqa} for reviews). Provided that these symmetries are exact or preserved with high accuracy, the topological defects are stable and, once formed, can persist in the universe for a long  time, thereby influencing the subsequent evolution of the universe. Topological defects play an important role in extracting information about physics beyond the Standard Model from cosmological observations, as they carry information about UV theory at extremely high energies. In particular, gravitational waves (GWs) are known to be emitted by the evolution and collapse of topological defects whose energy is spatially localized.

In this Letter, we study GWs generated by domain walls (DWs) - created by spontaneous breaking of discrete symmetries - in scenarios where they are destabilized and collapse by explicit breaking of discrete symmetries~\cite{Vilenkin:1981zs,Preskill:1991kd,Chang:1998tb,Gleiser:1998na,Hiramatsu:2010yz,Kawasaki:2011vv,Hiramatsu:2013qaa,Nakayama:2016gxi,Saikawa:2017hiv}. The GWs resulting from DW annihilation are characterized by their peak frequency and peak height. The peak frequency is determined by the timing of the DW annihilation, which can be used to extract the information of the UV physics~(see e.g., Refs.~\cite{Takahashi:2008mu,Dine:2010eb,Jaeckel:2016jlh,Higaki:2016yqk,Higaki:2016jjh,Chen:2021wcf,Ferreira:2022zzo,King:2023cgv, Madge:2023cak}). 
The peak height depends on the energy density of the DWs at the decay.

In most of the previous literature on the GWs generated by DW annihilation, the tension and the potential bias are treated as independent parameters, and in particular, the bias is often assumed to be a constant parameter. 
In this case, the peak frequency is determined by the balance between the tension and the potential bias. 
Instead, we propose here a scenario in which the DWs annihilate at a more or less fixed time, thanks to the temperature-dependent potential bias, which gives a more definite prediction about the peak frequency of the produced GWs. In particular, while the majority of previous studies have considered a constant bias in the context of GWs from DW annihilation, none have incorporated any form of bias into their numerical lattice calculations. Our research breaks new ground by introducing this temperature-dependent bias into lattice simulations for the first time.

As a candidate for DWs, we consider axion DWs. Axions, whose existence is strongly suggested by string theory, have attracted much attention in recent years and have been actively studied both theoretically and experimentally in various contexts, including dark matter, dark energy, and inflation. One of the features of axions is the periodicity of the potential, which naturally includes DW configurations.

We focus on the evolution of axion DWs when the axion is coupled to Standard Model gauge bosons, in particular gluons. 
In this case, due to the non-perturbative effects of QCD, a temperature-dependent explicit symmetry breaking is expected to grow as the temperature of the universe approaches the QCD scale, $T_{\rm QCD} \sim 100$\,MeV, leading to the decay of DWs around that time.  
While the potential bias is usually assumed to be a constant, in our scenario the potential bias is negligibly small well above the QCD scale, but rapidly increases as the temperature approaches $T_{\rm QCD}$. 
Thus, for the DW tension in a certain range, the DWs necessarily collapse near the QCD scale temperature, $T_{\rm QCD}$. 
Therefore, this scenario predicts a very specific frequency range of GWs associated with the DW annihilation, which falls in the range of about ${\cal O}(1-10)$\,nHz~\cite{Higaki:2016jjh,Ferreira:2022zzo}. 
This is within the sensitivity range of the pulsar timing array. 

Very recently, the NANOGrav, EPTA, PPTA, and CPTA groups published data showing the detection of stochastic GW background in the ${\cal O}(1-10)$\,nHz frequency band~\cite{NANOGrav:2023gor,Antoniadis:2023ott,Reardon:2023gzh,Xu:2023wog}.
Supermassive black holes are a strong candidate for the origin of the GWs in this frequency range, and other new physics explanations are also possible such as the first-order phase transition in a hidden sector~\cite{Nakai:2020oit,Ratzinger:2020koh} (see Ref.~\cite{NANOGrav:2023hvm} and references therein for other possibilities). 
Among various possibilities, the above scenario that the QCD crossover triggered the collapse of axion DWs is very attractive because it naturally explains the detected frequency range~\cite{Higaki:2016jjh,Ferreira:2022zzo}. We will see that the predicted GW spectrum agrees very well with the recently published pulsar timing data~\cite{NANOGrav:2023gor,Antoniadis:2023ott,Reardon:2023gzh}.

While previous numerical analyses of GWs from DW collapse have generated fitting models, they have not incorporated a potential bias into the lattice calculations~\cite{Hiramatsu:2010yz,Hiramatsu:2013qaa}. This bias has often been used simply as a timing parameter for collapse, measured at the point where the DW tension equals the bias. This widely accepted model, however, ignores the fact that  DW collapse extends over a period of time. This simplification was introduced because introducing a constant bias would cause the DW to begin to collapse before approaching the scaling solution. We note that Refs.~\cite{Kawasaki:2014sqa,Saikawa:2017hiv} suggested DW annihilation later than the above na\"{i}ve estimate, although they did not perform lattice calculations for the associated GW estimates.

In contrast, our approach considers a temperature-dependent bias that is initially negligible. This allows the DW to collapse within the numerical time frame after the scaling solution is realized. In a pioneering effort, we incorporate this bias into our numerical lattice calculations, allowing for a more thorough assessment of GW generation throughout the DW decay process. This methodology paves the way for accurate GW spectral estimates, bridging observed pulsar timing data with theoretical predictions. The primary goal of this Letter is to highlight these results. Our results suggest that late-stage DW collisions, significantly driven by the false vacuum energy, are the main contributors to GWs, a facet previously overlooked in the extrapolations of GWs in the scaling regime.

We comment here on the relevant literature in the past. 
Axion DWs or string-wall networks that collapse due to the potential bias arising from the non-perturbative QCD effect were considered in Refs.~\cite{Higaki:2016jjh,Higaki:2016yqk}.
In particular, it was pointed out in Ref.~\cite{Higaki:2016jjh} that the annihilation of the axion string-DW network produces GWs in the nanohertz range. 
The setup was motivated by the so-called clockwork QCD axion model~\cite{Higaki:2015jag}, where there are several heavy axions coupled to QCD. 
These heavy axions can be searched for in accelerator experiments~\cite{Higaki:2016jjh,Higaki:2016yqk}. 
A similar scenario with axion DWs coupled to QCD was also recently discussed in Ref.~\cite{Ferreira:2022zzo}, where the implications for the observed data of the pulsar timing array, as well as various collider and beam-dump experiments, were studied in detail.

{\bf Axion domain walls.--} 
Let us introduce the axion, $\f$, as a 
field that satisfies the periodic condition, 
\begin{equation}
\laq{sym}
\phi \;\to\; \phi + 2\pi f_\phi,
\end{equation}
where $f_\phi$ is the decay constant.  The action is exactly invariant under this translation. 
Since the potential also remains invariant under the shift, we  have degenerate vacua at $\f=\vev{\f} $ and $\f=\vev{\f} + 2\pi f_\f.$ 
So there is a DW configuration separating the two adjacent vacua.
For instance, we can consider 
 the axion potential of the simple cosine form,
\begin{equation}
\label{vphi}
V(\phi)=\frac{m_\phi^2 f_\phi^2}{n^2} \(1-\cos{\(n\frac{\phi}{f_\f}\)}\) 
\end{equation}
where $n$ is a non-zero integer to satisfy \Eq{sym}.
Here $m_\phi$ is the mass of the axion, and we assume $m_\phi \ll f_\phi$. 
Such a potential can be generated by some non-perturbative effects.
We introduce a coupling to QCD below. While this coupling also contributes to the mass, its effect is negligible and will be neglected in the following discussions

This potential has degenerate minima at $\phi =  2 \pi m f_\phi/n$ where $m$ is an integer.
Thus, we have DW configurations where $\phi$ changes from $2\pi m f_\f/n$ to $2\pi (m+1)f_\f/n$.
Each DW has a tension of
\begin{equation}
\s=  \frac{8}{n^2}  m_\f f_\f^2 .
\end{equation}
If we embed the axion in a phase of a complex scalar field as $\Phi = f_\phi/\sqrt{2} e^{i\phi/f_\phi}$, $n$ is identified with the so-called DW number, which counts the number of physically distinct vacua. 
The shift symmetry of the axion is now identified with the Peccei-Quinn $\U(1)_{\rm PQ}$ symmetry of $\Phi$.
It is possible that at some point in the history of the universe, $\Phi$ develops a nonzero vacuum expectation value, spontaneously breaking the $\U(1)_{\rm PQ}$ symmetry. Cosmic strings are then formed. The axion potential (\ref{vphi}) further breaks it down to the $Z_n$ subgroup, leading to the formation of DWs.
If $n = 1$, a DW is attached to each axion string, and the string-wall system becomes unstable due to the tension of the  DWs, and soon disappears after the formation of the DWs.
On the other hand, if $n \geq 2$, the string-wall system  is stable.

Even if $n \geq 2$, the string-wall system becomes unstable if the degeneracy of the vacua is lifted by other contributions to the potential.
One possible contribution to the potential could come from the interaction with the Standard Model particles.
In particular, $\phi$ can couple to gluons through
\begin{equation}
{\cal L}\supset -\frac{\a_s}{8\pi} \left(n_{g}\frac{\f}{f_\f}+\theta \right)G_{\mu \nu}^a\tl{G}^{\mu \nu}_a,
\end{equation}
where $\a_s$ is the strong coupling constant,
$G$ the gluon field strength, and $\tl{G}$ its dual. Here $n_{g}$ is another integer due to \Eq{sym}, and $\theta$ is the strong CP phase. 
Through this coupling, the axion acquires the QCD-induced potential,
\begin{align}
    V_{\rm QCD}(\phi)
    &=
    \chi(T) \left[ 
        1 - \cos \left( n_{g}\frac{\f}{f_\f}+\theta \right)    
    \right] ,
\end{align}
where the topological susceptibility is given by
\begin{align}
    \chi(T)
    =
    \left\{
        \begin{array}{ll}
            \chi_0 & \quad (T < T_\mathrm{QCD})
            \\
            \chi_0 \left( \frac{T}{T_\mathrm{QCD}} \right)^{-c} & \quad (T \geq T_\mathrm{QCD})
        \end{array}
    \right.
    \ .
\end{align} 
Here, we adopt $c = 8.16$~\cite{Borsanyi:2016ksw}, $\chi_0 = (75.6\,\mathrm{MeV})^4$, and $T_\mathrm{QCD} = 153\,\mathrm{MeV}$.
For instance with $n_{g}=1$ and $\theta=0$, $V_\mathrm{QCD}$ has a true minimum at $\f= 0$.
For $n_{g}\neq 1$, there are other minima if $n_{g}$ and $n$ have a common divisor other than unity.

\begin{figure}[!t]
    \begin{center}  
    \includegraphics[width=80mm]{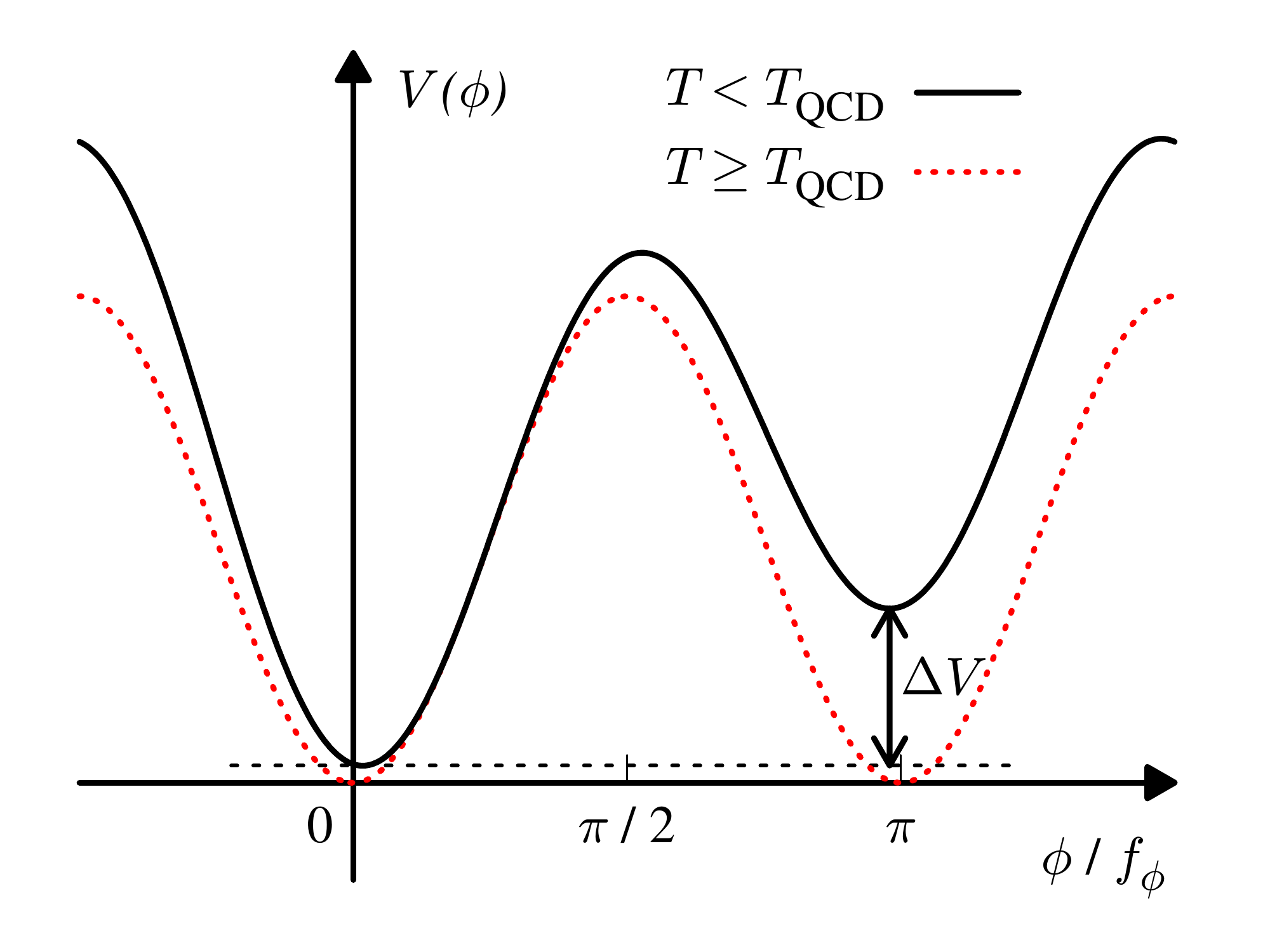}
    \end{center}
    \caption{The potential shape for $n_{g}=1 \AND n=2$. The red dotted and black solid lines denote the case with $T\gg T_{\rm QCD}$, and $T\leq T_{\rm QCD}$, respectively.}
    \label{fig:pote} 
\end{figure}

We consider a situation where some stable DW configurations, present at $T\gg T_{\rm QCD}$, become unstable and disappear at $T \sim T_{\rm QCD}$ due to the potential bias $V_{\rm QCD}$.
In the following, we will mainly focus on 
the case of
\begin{equation} 
n=2, \quad n_{g}=1 ,
\end{equation}
for simplicity of discussion. 
The potential shape is shown in Fig.\,\ref{fig:pote},
where 
the size of the bias is exaggerated for illustration purpose.
While the potential has degenerate minima at $T \gg T_\mathrm{QCD}$, the degeneracy is lifted by $V_\mathrm{QCD}$ at $T \lesssim T_\mathrm{QCD}$.
The difference in the potential at two local minima, or simply the bias, varies from $0$ to $2 \chi_0$ depending on $\theta$ in the case of $n = 2$ and $n_g = 1$, but its typical value is of the order of $\chi_0$.

{\bf GWs from collapse of DW network.--}
The DWs begin to annihilate when the bias becomes comparable to the DW energy density. 
Following the earlier studies, we 
define $T_{\rm eq}$ as the temperature at which the force applied by the tension equals the pressure from the potential bias. This can be represented as
\begin{equation} 
   \Delta V = {\cal F} \chi(T_{\rm eq}) 
    =
    H(T_{\rm eq}) \s,
    \label{eq: Tann defenition}
\end{equation}
 where $\Delta V$ is the potential bias, and 
${\cal F}$ is an $\mathcal{O}(1)$ constant accounting for the dependence on $\theta$ and $n_{g}$; {e.g., ${\cal F}=
 2\cos(\theta)$ for $n=2$ and $n_g=1$. {$H$ is the Hubble parameter}, and we assume radiation dominated Universe.}
Note that it takes some more time for most of the DWs to annihilate due to the bias~\cite{Kawasaki:2014sqa,Saikawa:2017hiv}.
We define the temperature at which most of the DWs annihilate by $T_\mathrm{ann}$, which is different from $T_{\rm eq}$ as we will see.

We note that, compared to the usual case with a constant bias,  the bias $\chi(T)$ depends on temperature.
This means that as long as ${\cal F} \chi(T_{\rm QCD}) \gtrsim H(T_{\rm QCD}) \s $, i.e.,
{$
\sigma{\cal F}^{-1}\lesssim  2\times 10^{15} \GEV^3,
$}
the DWs begin to annihilate around the QCD scale. 
 {Furthermore, the tension of the DWs cannot be much greater than this, since the DWs or the potential bias in the false vacuum would dominate the universe before the DWs are annihilated by the bias. In either case, this would lead to a significantly inhomogeneous Universe.
Specifically, we need $\s H \lesssim \Delta V$ when $\Delta V \sim 3H^2 M^2_{\rm pl}$ to avoid the cosmological catastrophe.
}
This leads to\footnote{
By requiring no false vacuum inflation occurs in any Hubble patch in the observable Universe, the condition becomes severer. Also,
from an anthropic point of view, it may be preferable for the tension to saturate the bound.
}  
$
\sigma {\cal F}^{-1} \lesssim 2\times 10^{16}\GEV^3.
$

{\bf Analytic estimate of the GW spectrum.--} 
Following Ref.~\cite{Hiramatsu:2013qaa}, we assume that the DWs instantaneously annihilate at $T = T_\mathrm{eq}$, i.e., $T_\mathrm{ann} = T_\mathrm{eq}$. Note, however, that our forthcoming numerical calculations will show that this assumption of the instantaneous annihilation is overly simplistic.

Even before the DWs annihilate, the DWs are continuously decaying in the scaling regime.
The energy of the decaying DWs is released to the axion particles and GWs.
The spectrum of the GWs emitted from the decaying DWs has a peak corresponding to the Hubble parameter at the DW decay~\cite{Hiramatsu:2013qaa}.
Since the energy fraction of the DWs increases in the scaling regime, the peak frequency of the current GW spectrum is determined by the Hubble parameter at the DW annihilation, 
\begin{equation}
    \laq{freq} 
    f_{\rm peak} 
    \equiv
    H(T_{\rm ann}) = H(T_{\rm eq}),
\end{equation}
where we used $T_\mathrm{ann} = T_\mathrm{eq}$.
Taking into account the redshift of the peak frequency due to the {subsequent} cosmic expansion, one obtains the present value of the peak frequency as 
\begin{align}
    \tilde{f}_{\rm peak,0} 
    \simeq 
    13 \, {\rm nHz}  \left(\frac{g_{\star}(T_{\rm eq})}{20}\right)^{1/6}\left(\frac{T_{\rm eq}}{0.1\,{\rm GeV}}\right) ,
\label{eq: f peak analytic}
\end{align}
{where the tilde represents the use of the assumption of $T_\mathrm{ann} = T_\mathrm{eq}$.}
Here, $g_{\star}(T)$ is the relativistic degree of freedom for the energy density. {The subscript} $``0"$  denotes the quantity in the present Universe here and hereafter. 

The peak value of $\Omega_{\rm GW}$ at the DW annihilation can be estimated as
$
    \tilde{\Omega}_{\rm GW,ann}^{\rm peak} 
    =
    \left.
        \frac{ \epsilon_{\rm GW}  \mathcal{A}^2 \sigma^2}{24 \pi M_{\rm pl}^4  H^2}
    \right|_{T=T_{\rm eq}},
$ 
where {$M_{\rm pl}\approx 2.4\times 10^{18}\,$GeV is the reduced Planck mass,} $\epsilon_{\rm GW} \simeq 0.7 \pm 0.4$~\cite{Hiramatsu:2013qaa}
parametrizes the efficiency of GW emission, and $\mathcal{A}$ is the so-called area parameter defined by $\rho_{\rm DW} = \mathcal{A} \sigma /t$. {Here $\rho_{\rm DW}$ is the DW energy density.}
When $T \simeq T_\mathrm{eq}$, the area parameter is $\mathcal{O}(1)$ and the bias is comparable to the DW energy density, i.e., $\rho_{\rm DW} \sim {\cal F} \chi(T_\mathrm{eq})$.
{Then we get 
\begin{equation}\laq{hight}
    \tilde{\Omega}_{\rm GW,ann}^{\rm peak} =  \left.\frac{\x^2 {\cal F}^2 \chi^2}{24 \pi M_{\rm pl}^4  H^4}\right|_{T=T_{\rm eq}}.
\end{equation}}%
Here, we have introduced a parameter $\x$ to include various numerical factors such as
the area parameter, and the efficiency of the GW emission,
{and we will take $\xi = 1$ for simplicity.}
The present value of $\Omega_{\rm GW}$ {at the peak frequency is given by}
\begin{align}
    \tilde{\Omega}_{\rm GW,0}^{\rm peak}h^2 
    =
    \Omega_{r}h^2
    \frac{g_{\star}(T_\mathrm{eq})}{g_{\star,0}}
    \left(
        \frac{ g_{\star s, 0} }{ g_{\star s}(T_{\rm eq}) }
    \right)^{4/3}
    \tilde{\Omega}_{\rm GW, ann}^{\rm peak}
    \ ,
\end{align}
where $\Omega_rh^2 = 4.15 \times 10^{-5}$ is the density parameter of radiation, and $g_{\star s}(T)$ is the relativistic degree of freedom for the entropy density. 
Thus, we obtain
\begin{align}
    \tilde{\Omega}_{\rm GW,0}^{\rm peak}h^2 
  \simeq      &
    2.7 \times 10^{-10} 
    \left( \frac{g_{\star s}(T_\mathrm{eq})}{20} \right)^{-\frac{4}{3}}
    \left( \frac{g_{\star}(T_\mathrm{eq})}{20} \right)^{-1}
    \x^2 {{\cal F}^2}\nonumber \\
    \times &
    \left\{
        \begin{array}{ll}
            \left( \frac{T_{\rm QCD}}{T_{\rm eq}} \right)^{8+2c}
            & \quad (T_{\rm eq} {\geq} T_\mathrm{QCD})
            \\
            \left( \frac{T_{\rm QCD}}{T_{\rm eq}} \right)^8 
            & \quad (T_{\rm eq} {<} T_\mathrm{QCD})
        \end{array}
          \label{eq: OmegaGW peak analytic}
    \right.
    \ ,
\end{align}
where we used $g_{\star,0} = 3.36$ and $g_{\star s, 0} = 3.91$.
{Numerical simulations showed that when compared to the peak frequency, the frequency dependence of $\Omega_\mathrm{GW,0}$ behaves as 
$\Omega_{\rm GW} \propto f^3$ for lower frequencies and  $\Omega_{\rm GW} \propto f^{-1}$ for higher frequencies~\cite{Hiramatsu:2013qaa}.}

{While  we have so far assumed the instantaneous DW annihilation at $T = T_\mathrm{eq}$, it is 
delayed compared to the balance between the bias and tension in realistic situations.
Two key factors influence the GW spectrum during this delay. Firstly, the peak frequency of the GW spectrum shifts to lower frequencies, enhancing the peak abundance. Secondly, and more critically, there is the addition of GWs generated during the DW annihilation -- a factor not considered in earlier literature. The spectrum of the latter GWs differs from those produced in the scaling regime. See Supplemental Material for details.}
 
We parameterize 
{the actual peak frequency and the peak abundance using factors, $C_f$ and $C_\Omega$:}
\begin{align}
    f_{\rm peak,0} 
    &=
    C_f \tilde{f}_{\mathrm{peak},0}
    .
    \label{eq: f peak realistic}
    \\
    \Omega_{\mathrm{GW},0}^{\mathrm{peak}}
    &=
    C_\Omega 
    \tilde{\Omega}_{\mathrm{GW},0}^{\mathrm{peak}}
    .
\end{align}
{The peak abundance is modified by the multiple effects; the delay of the annihilation and the boost by the bias enhance it, while the deviation from the scaling regime reduces it.}

{\bf {Numerical estimate of the GW spectrum.}--}
We have performed the lattice calculation to numerically simulate the annihilation of the DWs with a {temperature-dependent} bias and evaluated the GW spectrum based on the analysis in Ref.~\cite{Lee}.%
We show the {current} GW spectrum from the DW annihilations for 
{different tensions}
in Fig.~\ref{fig: GW spectrum}, where we have set {$\mathcal{F} = 1$ and} $g_\star = g_{\star s} = 10.75$
and $20$.%
\footnote{
{We adopt the values of $g_\star$ and $g_{\star s}$ since the DW annihilation occurs over a duration and the numerical simulation does not consider the time-dependence of $g_{\star}$ and $g_{ \star s}$. The region between the solid and dashed lines accounts for the uncertainty in the matching.  }
}
The lines with the same color share the same numerical setup and correspond to different $\sigma$ because the {matching} between the numerical and physical parameters depends on $g_*(T_\mathrm{eq})$.
For all cases, we obtain $C_f < 0.5$ and $C_\Omega > 10$.
For example, we find $C_f \approx 0.47 
$ and $C_\Omega \approx 20
$ for the green solid line. 
See Supplemental Materials
for more details on the matching method.
We find that the obtained spectrum {agrees} well with
{the NANOGrav 15\,yr result~\cite{NANOGrav:2023hvm} for $\sigma \approx 10^{15}\,\mathrm{GeV}^3$.}
Since the GW strains reported by the EPTA~\cite{Antoniadis:2023ott}, PPTA~\cite{Reardon:2023gzh}, and CPTA~\cite{Xu:2023wog} groups show a similar trend to that of the NANOGrav, the GW spectrum in our scenario {is also consistent} with them.
{Interestingly, the range of axion parameters that can explain the NANOGrav results is large, and axions in this range can be explored in accelerator experiments (see Supplemental Material).}

{In our numerical simulation, GWs are primarily produced by DWs that are accelerated by the bias during their collapse,  rather than the conventional ones from DWs in the scaling regime. 
This dominance is due to several interplaying factors: the GWs are enhanced as the timing of the DW collapse is delayed. At the same time, the bias accelerates the DWs, amplifying their energy. However, as the collapse progresses, the typical size and number of DWs decrease. This complex combination of effects leads to the observed GW enhancement.}

\begin{figure}[!t]
    \begin{center}  
        \includegraphics[width=85mm]{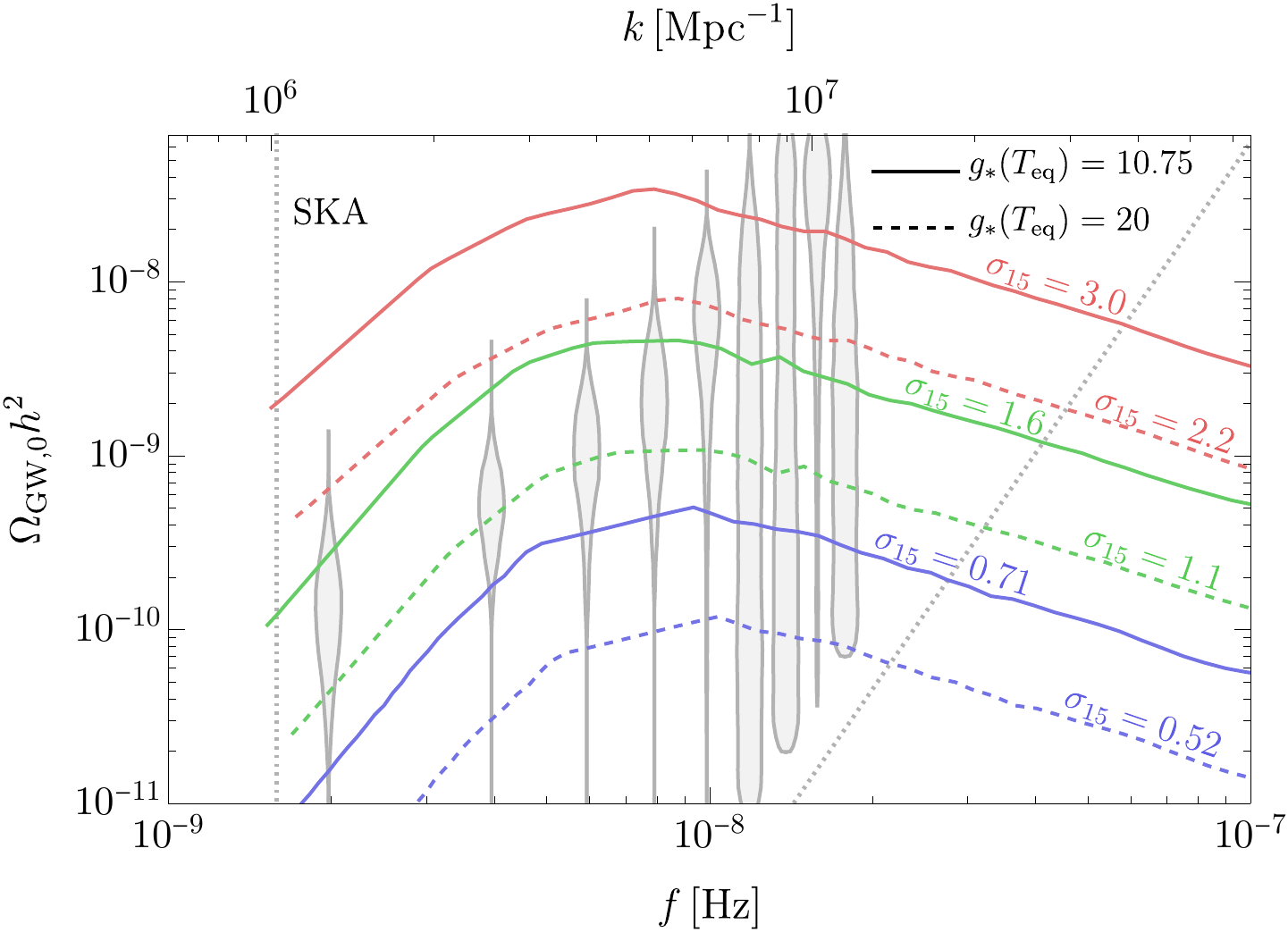}
    \end{center}
    \caption{%
        GW spectrum from the DW annihilations for several values of the tension, $\sigma_{15} \equiv \sigma/(10^{15}\,\mathrm{GeV}^3)$.
        The solid and dashed lines represent the results for $g_*(T_\mathrm{eq}) = 10.75$ and $20$, respectively.
        The gray violins are from the NANOGrav 15\,yr results~\cite{NANOGrav:2023hvm}.
        Here, we show nine bins of the data in the low-frequency range, where the GW abundance is well determined.
        The gray {dotted} line shows the future sensitivity of SKA~\cite{Janssen:2014dka}, evaluated based on Ref.~\cite{Inomata:2018epa}.
    }
    \label{fig: GW spectrum} 
\end{figure}

Recently a DW network with initial inflationary fluctuations was shown to be quite stable against the population bias~\cite{Gonzalez:2022mcx}. 
In these situations, the resultant GW spectrum is expected to resemble our current results, but will likely be broader. This may provide a better fit to the NANOGrav data or other PTA data.
 (See Ref.~\cite{Lee} for more details.)

{\bf Conclusions.--}
\label{sec: conclusions}
We have shown that the axion DWs coupled to QCD gluons can explain the recent data from NANOGrav, EPTA, PPTA, and CPTA. The DW bias emerges at the cosmic temperature near the QCD crossover, aligning well with the frequency observed in pulsar timing data. In estimating the GWs produced by  DW annihilation, 
we have performed the pioneering lattice simulations that incorporate a temperature-dependent bias for the first time. We have found that the GWs produced during DW collapse make the dominant contribution to the GW spectrum, and their contribution is different from that in the scaling regime.
Our scenario may be probed by searching for axions in future accelerator experiments. Additionally, the DW collapse might produce PBHs with about a solar mass, potentially accounting for LIGO/Virgo black hole merger events (see Supplemental Material).  

\section*{Acknowledgments}
This work is supported by JSPS Core-to-Core Program (grant number: JPJSCCA20200002) (F.T.),  JSPS KAKENHI Grant Numbers 23KJ0088 (K.M.),  20H01894 (F.T. and N.K.), 20H05851 (F.T., W.Y., and N.K.),  21K20364 (W.Y.),  22K14029 (W.Y.), and 22H01215 (W.Y.), 21H01078 (N.K.), 21KK0050 (N.K.), Graduate Program on Physics for the Universe (J.L.), and Watanuki International Scholarship Foundation (J.L.). This article is based upon work from COST Action COSMIC WISPers CA21106,  supported by COST (European Cooperation in Science and Technology). Part of the results in this Letter were obtained using supercomputing resources at Cyberscience Center, Tohoku University.
\appendix
\clearpage
\section*{Supplemental Material}
To supplement our main discussion, we outline the setup for the numerical simulation, illustrate the time evolution of the GW spectrum, consider solutions to the strong CP problem in line with our scenario, and highlight other phenomena pertinent to this scenario. 
\section{Numerical simulation}
Here, we explain the setup for the lattice simulation.
Our analysis here is based on Ref.~\cite{Lee}.
We employ a $4096^3$ lattice of the $\phi^4$ theory (see Refs.~\cite{Kitajima:2022jzz,Gonzalez:2022mcx} for the setup) with a box that initially contains $50^3$ Hubble horizons. The potential is given by
\begin{equation}
V(\phi) = V_0-\frac{1}{2}m_0^2 \phi^2 + \frac{\lambda}{4}  \phi^4.
\end{equation}
The vacuum expectation value, $v$, of the scalar is determined by $m_0^2=\l v^2$. The scale factor $a(\tau)$ is set to unity when $H=m_0$, and we assume the radiation dominated universe.
Here the conformal time $\tau$ is normalized  so that 
the Hubble parameter is given by $H(\tau) = (m_0 \tau^2)^{-1}$. In other words,
$\tau = 1/m_0$  at the initial time when $H = m_0$.

To replicate the behavior of the axion potential resulting from QCD, we introduce a time-dependent potential bias as
\begin{align} 
{\rm V}_{\rm bias}(\phi,\tau) &=  \lambda v^3 \phi \times b(\tau) ,
\\ 
b(\tau)&\equiv\frac{\epsilon }{1+e^{-2(\tau - \tau')/\delta \tau}}
\end{align}
where $\tau' = 4/m_0$ {and} $\delta \tau = 0.2/m_0$.
The magnitude of the bias is determined by $\epsilon$. We investigate various values for $\epsilon$, specifically, $\epsilon = 0.025, 0.05$, and $0.1$. Note that the bias has a different form from the realistic QCD-induced potential for alleviating the numerical cost, but we expect that this difference does not significantly change our results.
The conformal time $\tau_\mathrm{eq}$ corresponding to the temperature $T_{\rm eq}$ (defined in the main text) is obtained  by solving
$\Delta V = 2\lambda v^4  b(\tau_\mathrm{eq}) = 2\sqrt{2} v^2/(3 \tau_\mathrm{eq}^2)$.
Here, we have used $\sigma = 2\sqrt{2} m_0 v^2/3$ in the $\phi^4$ model.

In Fig.~\ref{fig:lattice}, 
we present the numerical results of the GW spectra for the case of $\epsilon = 0.025,\ 0.05$, and $0.1$.
We have followed the evolution of the system over the conformal time $\tau$ in the range of $(1\,\text{--}\,80)/m_0$ and confirmed that the DWs entirely disappear by $\tau\sim 20/m_0$. Beyond this point, the GW spectrum ceases to evolve.
For the range $\tau = 5/m_0  - 15/m_0$, spectra are shown at intervals of $\Delta \tau = 1/m_0$, and from $\tau = 20/m_0$ to $\tau = 80/m_0$, they are shown at intervals of $\Delta \tau = 10/m_0$. Among them, the black dashed 
line corresponds to $\tau = 10/m_0$,
while the black solid line represents $\tau = 80/m_0$. Here the density parameter $\Omega_{\rm GW}$ is evaluated at the time of the production, and it is a combination of $\Omega_{\rm GW}(k) (M_{\rm pl}/v)^4$ that is calculated in the numerical simulations. {$k$ is the wavenumber evaluated on the comoving coordinates.} To obtain the current GW spectrum, 
{one needs} to determine the physical value of $v$ using the matching method we will describe shortly, and multiply the dilution factor due to the redshift as well.

Note that, while DWs are collapsing due to potential bias, they are boosted by the false vacuum energy, and acquire a typical spatial scale smaller than in the scaling regime.
{Focusing on the case of $\epsilon=0.05$} {(the middle figure)},  one can clearly see that for $\tau \approx 10/m_0$ (indicated by {the black dashed line}), a dominant source of GW emerges from the subhorizon scale around $k/m_0 \sim 1$, whereas the peak in the scaling regime at that time corresponds to $k/m_0 \sim 0.5$. This contribution at subhorizon scales  comes to dominate the whole GW spectrum afterwards. This is attributed to the collapse of the DWs accelerated by the potential bias, which has not been considered in earlier studies~\cite{Hiramatsu:2010yz,Kawasaki:2011vv,Hiramatsu:2013qaa}. 
For the case of $\epsilon = 0.025 ~({\rm and}~ 0.1)$, the subhorizon peak rises slowly (quickly), which is also consistent when it comes from collapse of boosted DWs.

To translate the numerical results to the physical ones, we match quantities at $\tau = \tau_\mathrm{eq}$. 
Specifically, we first determine $T_\mathrm{eq}$ by matching the relative height of the bias as $b(\tau_\mathrm{eq})/\epsilon = \chi(T_\mathrm{eq})/\chi_0$.
Then, we solve $H(T_\mathrm{eq}) = (m_0 \tau_\mathrm{eq}^2)^{-1}$ and $\chi(T_\mathrm{eq}) = 2\lambda v^4  b(\tau_\mathrm{eq})$ to determine the corresponding physical values of $\lambda$ and $v$, where we set $\mathcal{F} = 1$ for simplicity. Here $H$ is the physical Hubble parameter, and we need to fix the value of $g_*(T_{\rm eq})$ to evaluate it. Since the time-dependence of $g_*$ is not included in the numerical simulation, we have adopted two different values $g_*(T_{\rm eq}) = 10.75$ and $20$.  This results in a slightly different estimate of $\Omega_{\rm GW}$ as shown in the main text, which should be regarded as the uncertainty of the matching.
One can also match the comoving wavenumber $k$ corresponding to the horizon scale at $T = T_\mathrm{eq}$. Then,
the physical GW spectrum is obtained by multiplying $(v/M_{\rm pl})^4$ to the one obtained in the numerical simulations. The dilution factor is taken into account by multiplying the radiation density parameter, except for a slight change in the relativistic degrees of freedom. 
We repeat the numerical simulations and the matching procedure for the case of $\epsilon=0.1 \AND 0.025$.
For each case, we determine $\sigma$ through the matching.
By identifying the peak location where the GW abundance is the maximum, we can evaluate $C_f$ and $C_\Omega$.
A summary of the matching of the parameters can be found in Table.~\ref{tab: parameters}.
See Fig.~2 in the main text for the current GW spectrum
for the three different biases.

\begin{table}[tbp]
    \centering
    \caption{Mapping from the numerical parameters to the physical ones.
    }
    \label{tab: parameters}
    \begin{tabular}{c c c c c c}
        $\epsilon$ & $T_\mathrm{eq}/T_\mathrm{QCD}$ & $C_f$ & $C_\Omega$ & $g_*(T_\mathrm{eq})$ & $\sigma~[\mathrm{GeV}^3]$ 
        \\
        \hline
        \multirow{2}{*}{0.025} & \multirow{2}{*}{1.003} & \multirow{2}{*}{0.45} & \multirow{2}{*}{31} & 10.75 & $3.0 \times 10^{15}$ 
        \\
        & & & & 20 & $2.2 \times 10^{15}$ 
        \\
        \multirow{2}{*}{0.05} & \multirow{2}{*}{1.07} & \multirow{2}{*}{0.47} & \multirow{2}{*}{20} & 10.75 & $1.6 \times 10^{15}$ 
        \\
        & & & & 20 & $1.1 \times 10^{15}$ 
        \\
        \multirow{2}{*}{0.1} & \multirow{2}{*}{1.16} & \multirow{2}{*}{0.46} & \multirow{2}{*}{15} & 10.75 & $7.1 \times 10^{14}$
        \\
        & & & & 20 & $5.2 \times 10^{14}$
        
    \end{tabular}
\end{table}

We note that in our analysis we did not include the axion strings to simplify the analysis. We believe that this is a fairly good approximation when $N_{\rm DW}=2$ because two DWs attached to a single string can be approximated by a single DW.

\begin{figure}[!t]
    \begin{center}  
         \includegraphics[width=80mm]{
         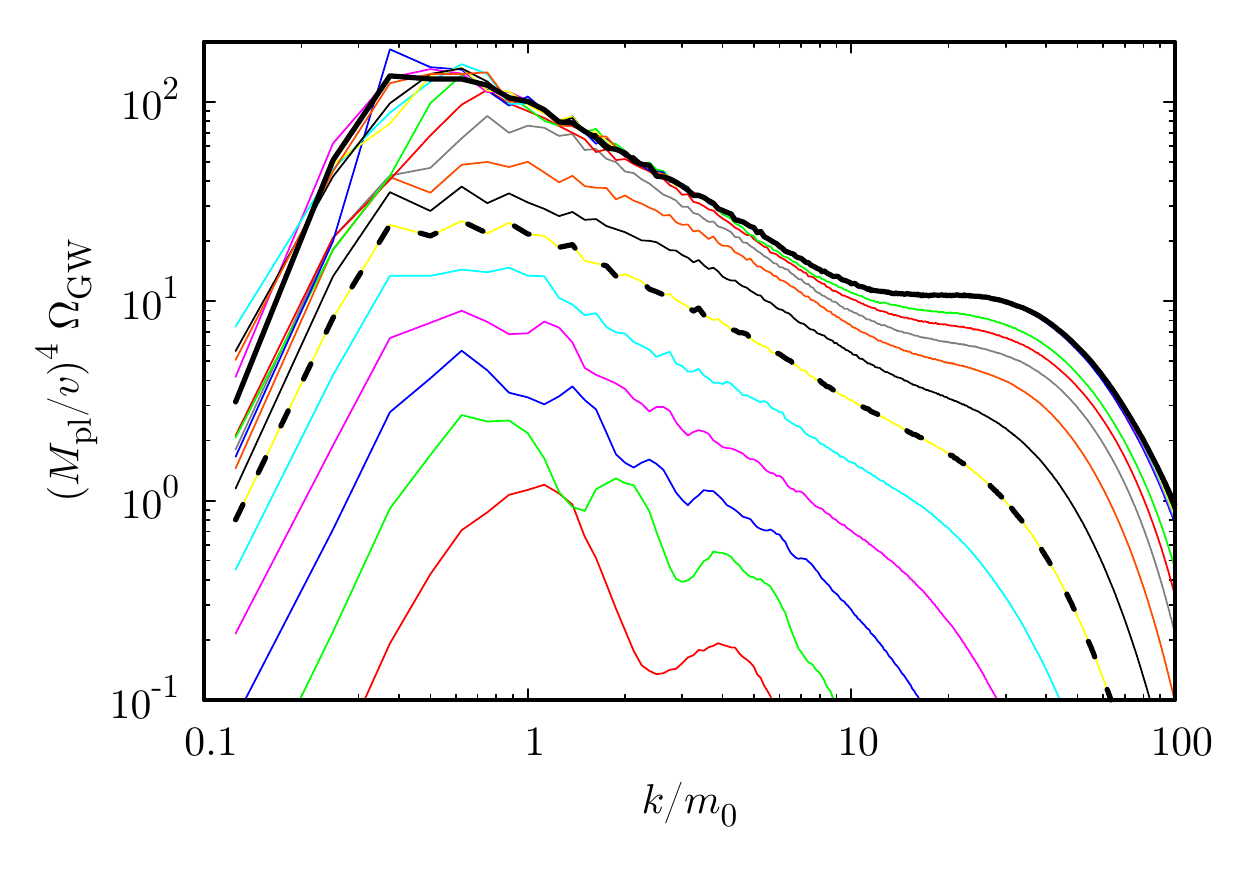
         }
         \includegraphics[width=80mm]{
         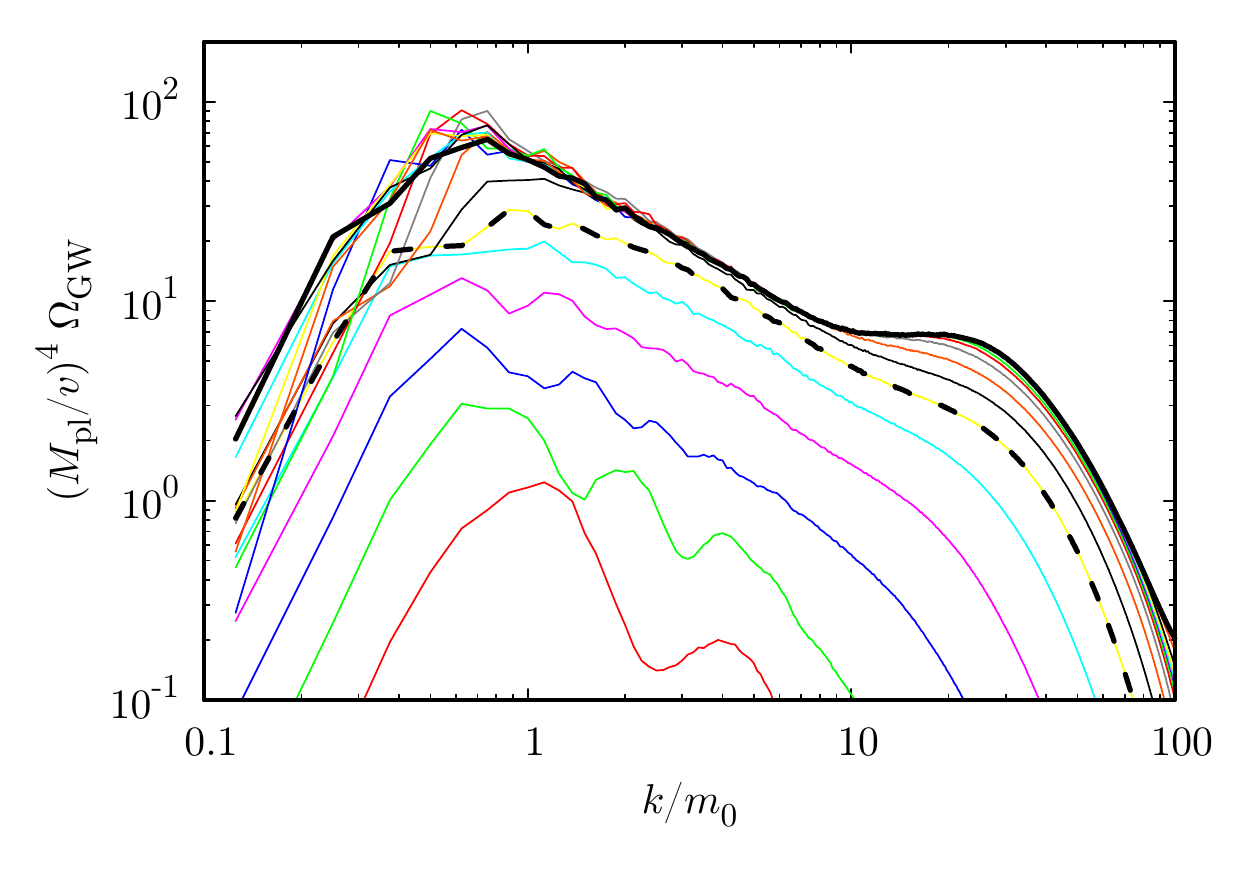
         }
         \includegraphics[width=80mm]{
         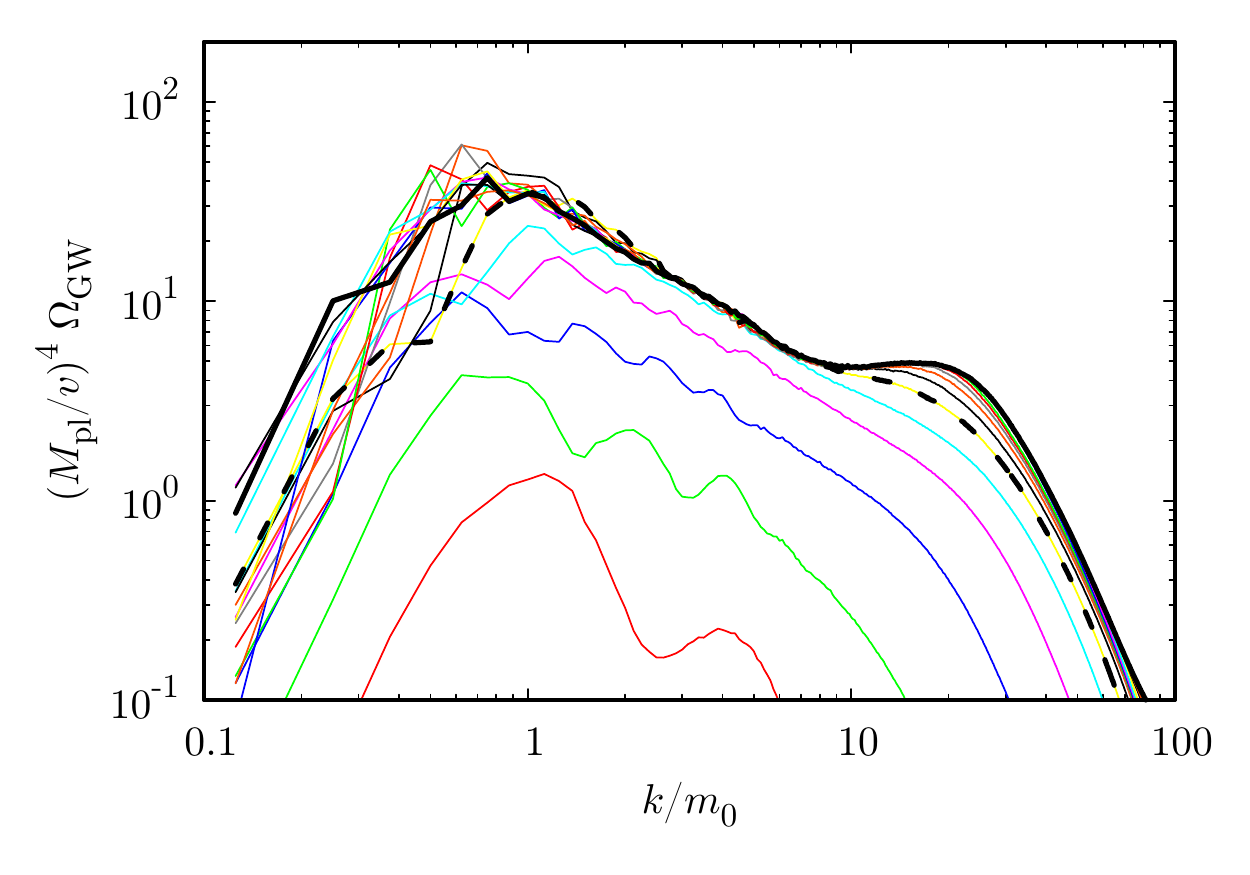
         }
    \end{center}
    \caption{
        Numerical results of the GW spectrum from the DW annihilations
        for $\tau = 5/m_0 \,\text{--}\, 80/m_0$ 
        with $\epsilon=0.025$ (top), $\epsilon=0.05$ (middle), $\epsilon=0.1$ (bottom).
        For $\tau \leq 15/m_0$,
        results are shown at intervals $\Delta \tau=1/m_0$, and for $20/m_0 \leq \tau \leq 80/m_0$, results are shown at intervals $\Delta \tau=10/m_0$. 
        The black dashed and lines represent the spectrum at the time $\tau = 10/m_0$ and $80/m_0$, respectively.}
    \label{fig:lattice} 
\end{figure}

\section{Solutions to Strong CP problem}
In this section, we address the strong CP problem within the context of our model. 
One approach is to ensure the strong CP phase is suppressed, specifically by having $|\vev{\f}/f_\phi+\theta|\ll 10^{-10}$, which satisfies the neutron electric dipole moment constraint. By considering the dynamics of a heavy QCD axion and its extended models,  the potential $V(\phi)$ could emerge from some UV physics that contains QCD, then the potential minima of $V(\phi)$ and $V_{\rm QCD}$ could be aligned.
Alternatively, the alignment between these two potentials could be attributed to inflationary dynamics, as discussed in Ref.~\cite{Takahashi:2021tff} (see also Ref.~\cite{Takahashi:2023vhv}).

We might also consider the presence of the light QCD axion to drive the strong CP phase to zero following the annihilation of DWs. This scenario is plausible if the QCD axion is sufficiently light, specifically when the decay constant $f_a\gtrsim 10^{17}\GEV$ (see Refs.\,\cite{Graham:2018jyp,Takahashi:2018tdu} for a detailed explanation on the mechanism to obtain the right abundance of axion dark matter without a fine-tuning of the initial misalignment angle).
If the QCD axion is relatively heavy and begins its oscillations before the DW annihilation, the evolution becomes complicated. The QCD axion would oscillate around different vacua in different regions. 
In this case, the primordial black holes (PBHs) or mini-clusters could be formed, which is analogous to the scenario of the so-called the QCD axion bubbles~\cite{Kitajima:2020kig}. In our model, for the DW to annihilate due to the potential bias generated by QCD, it is crucial that the QCD axion starts oscillating only after the DW annihilation has started. Otherwise, the oscillations of the QCD axion would reduce the effective size of the potential bias. In essence, the potential bias attributed to QCD is transformed into the energy of the QCD axion oscillation. Consequently, as the universe expands, the effective strength of the potential bias decreases progressively.

{With $f_a\ll 10^{12}\,\GEV$ one can delay the onset of the oscillation of the QCD axion until most of the DWs have annihilated. Such a delay in the onset of oscillation is required for the QCD axion to be the dominant dark matter from the misalignment mechanism.  One way of realizing this is to initially place the QCD axion on the hilltop of the axion potential in the false vacuum of the domains.  This hilltop axion can be realized by inflationary dynamics~\cite{Takahashi:2019pqf,Nakagawa:2020eeg,Narita:2023naj}.  In this case, the axion begins to oscillate as the DW passes through, which may provide a new mechanism for QCD axion dark matter with a decay constant smaller than $10^{12}$\,GeV.  Alternatively, one can consider that the axion is trapped via an explicit breaking term of the Peccei-Quinn symmetry and induces a trapped misalignment mechanism~\cite{Nakagawa:2020zjr,Jeong:2022kdr,Nakagawa:2022wwm}, realizing a late-time oscillation and axion dark matter. Depending on the potential form for the trapping, one can also end the trapping to have the QCD axion dark matter.} 

\section{Probing the Scenario: Other Relevant Phenomena}

The axion parameter region corresponding to the nHz GW background,
as suggested by the NANOGrav experiment, is shown in Fig.~\ref{fig:2} on the $m_\f$\,--\,$f_\f^{-1}$ plane. Here we show the parameter range of $\s\sim 10^{14-16}\GEV^3$, which is consistent with the NANOGrav results.
In our model, it is essential that the axion couples to gluons. We integrated constraints and prospective reaches from the paper~\cite{Takahashi:2021tff}, which adapted those from Ref.~\cite{Kelly:2020dda}. Moreover, considering the big-bang nucleosynthesis (BBN) bound with $m_\phi>0.2\GEV$, we have conservatively set the axion lifetime to be shorter than $0.01\,$sec. Note that
the spectrum of the axions produced from the DW annihilation is different
from those produced thermally (we also have thermally produced component, see e.g., Ref.~\cite{Depta:2020wmr} for the bound in this case).
Note that the viable parameter space extends to
 $m_\f \sim f_\f$, beyond which the perturbativity breaks down.
One can see that there is a large parameter space that is consistent with the NANOGrav data, which will be explored in upcoming or recent accelerator experiments. 

\begin{figure}[!t]
\vspace{20pt}
     \includegraphics
[width=80mm]{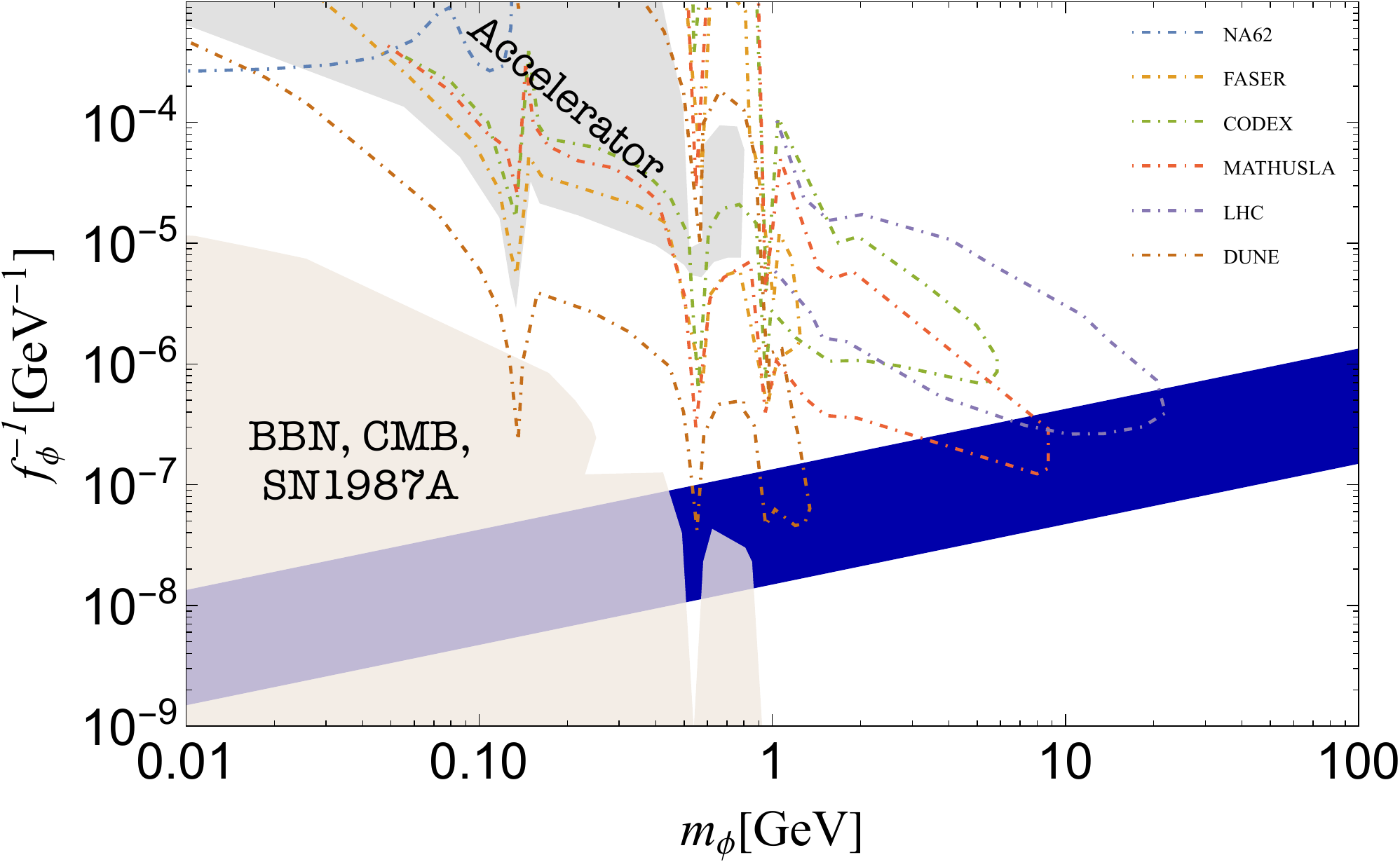}
\caption{The parameter range predicted by our scenario that can explain the GW signals suggested by the NANOGrav experiment is illustrated in the $m_\f$ -- $f^{-1}_\f$ plane within the blue band, considering $n_{g}=1, n=2$.
 The various bounds and future sensitivities are adopted from Ref.\,\cite{Takahashi:2021tff}. We incorporate expected reaches from experiments such as CODEX, DUNE, FASER, LHC, MATHUSLA, and NA62~\cite{Chou:2016lxi,Gligorov:2017nwh,Feng:2018noy,Ariga:2018uku,Hook:2019qoh,Aielli:2019ivi,Kelly:2020dda}, and from cosmological and astrophysical bounds~\cite{Vysotsky:1978dc,Raffelt:2006cw,Cadamuro:2011fd,Millea:2015qra,Zyla:2020zbs}. See also Ref.~\cite{Agrawal:2021dbo} for additional information on the theoretical uncertainties of the estimated experimental reaches. The BBN bound for $m_\f>0.2\GEV$ is set by requiring the lifetime of the axion to be shorter than $0.01$\,sec.
}
\label{fig:2} 
\end{figure}

In our scenario, PBHs may be formed around the time of DW annihilation.
This is because the vacuum energy in the false vacuum is not significantly smaller than the radiation energy density, which dominates the Universe.
When the size of such a false vacuum becomes comparable to the Schwarzschild radius, it could potentially collapse to form a PBH (see e.g., Ref.~\cite{Ferrer:2018uiu}). 
Comprehensive analysis, considering the volume fraction of the false vacuum during later stages $T\ll T_{\rm eq}$, the DW configurations, and angular momenta etc., will be crucial to accurately determine the PBH abundance.

On the other hand, we also expect that the PBH formation would be more likely when DWs are associated with initial inflationary fluctuations. In this case, the DW network has extensive voids as shown in Ref.~\cite{Gonzalez:2022mcx}. These sizable voids lead to large false vacuum regions, which collapse after a long time due to the potential and population bias, and form PBHs~\cite{Lee}. In such scenarios, it is important to consider the isocurvature constraint, because the DW network exhibits an almost scale-invariant power spectrum~\cite{Gonzalez:2022mcx} (See also Ref.~\cite{Lee}), and the resultant PBHs share this characteristic as well.

One of the robust predictions of our scenario is that, if PBHs are formed from the DW collapse, the PBH mass should be around $M_\odot$ since DWs annihilate around the QCD scale.
These PBHs may explain some or all of the LIGO/Virgo events~\cite{LIGOScientific:2016aoc,LIGOScientific:2016sjg,LIGOScientific:2017bnn,LIGOScientific:2017ycc} (see also Refs.~\cite{Bird:2016dcv, Clesse:2016vqa, Sasaki:2016jop}).


\bibliographystyle{apsrev4-1}
\bibliography{Ref}

\end{document}